\documentclass[onecolumn]{svjour3}                     
\smartqed  
\begin{document}
\title{ALMA Capabilities for Observations of Spectral Line Emission}
\titlerunning{ALMA Spectral Line Capabilities}
\author{Alwyn Wootten}
%
%
\institute{National Radio Astronomy Observatory (NRAO), 520 Edgemont Rd., Charlottesville, Virginia 22903, USA\\
              \email{awootten@nrao.edu}           
}

\date{Received: date / Accepted: date}

\maketitle

\begin{abstract}
The Atacama Large Millimeter/submillimeter Array (ALMA\footnote{The Enhanced Atacama Large Millimeter/submillimeter Array (known as ALMA) is an 
international astronomy facility. ALMA is a partnership between North America, Europe, 
and Japan/Taiwan, in cooperation with the Republic of Chile, and is funded in Europe by 
the European Southern Observatory (ESO) and Spain, in North America by the U.S. 
National Science Foundation (NSF) in cooperation with the National Research Council of 
Canada (NRC), and in Japan by the National Institutes of Natural Sciences (NINS) in 
cooperation with the Academia Sinica in Taiwan.   ALMA construction and operations 
are led on behalf of Japan/Taiwan by the National Astronomical Observatory of Japan 
(NAOJ), on behalf of North America by the National Radio Astronomy Observatory 
(NRAO), which is managed by Associated Universities, Inc. (AUI), and on behalf of 
Europe by ESO.}) combines large collecting area and location on a high dry site to provide it with unparalleled potential for sensitive millimeter/submillimeter spectral line observations. Its wide frequency coverage, superb receivers and flexible spectrometer will ensure that its potential is met. Since the 1999 meeting on ALMA Science\cite{RefA}, the ALMA team has substantially enhanced its capability for line observations. ALMA's sensitivity increased when Japan joined the project, bringing the 16 antennas of the Atacama Compcat Array (ACA), equivalent to eight additional 12m telescopes. The first four receiver cartridges for the baseline ALMA (Japan's entry has brought two additional bands to ALMA's receiver retinue) have been accepted, with performance above the already-challenging specifications. ALMA's flexibility has increased with the enhancement of the baseline correlator with additional channels and flexibility, and with the addition of a separate correlator for the ACA. As an example of the increased flexibility, ALMA is now capable of multi-spectral-region and multi-resolution modes. With the former, one might observe e.g. four separate transitions anywhere within a 2 GHz band with a high resolution bandwidth. With the latter, one might simultaneously observe with low spectral resolution over a wide bandwidth and with high spectral resolution over a narrow bandwidth; this mode could be useful for observations of pressure-broadened lines with narrow cores, for example. Several science examples illustrate ALMA's potential for transforming millimeter and submillimeter astronomy.
\end{abstract}

\section{Introduction: A Boom Time for IS Spectroscopy}
\label{intro}
Correlator technology has benefitted from huge increases in data processing ability in recent years.  The first of the new generation of correlators is already producing a flood of new spectral line data.

At NRAO, the Green Bank Telescope (GBT) combines a collecting an effective area roughly commensurate with ALMA's with powerful correlator capacity.  Furthermore, it provides nearly complete frequency coverage below 50 GHz, with initial bolometric array tests at 90 GHz producing promising results.  Nine new molecules have been identified in the past two years, including most recently the first negatively charged interstellar molecule, C$_6$H$^-$\cite{RefB}.  Enhancement of the Very Large Array (VLA) will provide it with vastly increased correlator capacity within the next few years, as well as nearly complete frequency coverage below 50 GHz.  Correlator capacity at IRAM and at CARMA has also recently increased and in the submillimeter regime, the Submillimeter Array is pioneering interferometric imaging in the last accessible atmospheric windows.


ALMA combine total power and interferometric modes of radio imaging.  It will provide complete frequency coverage, initially from 84-950 GHz, a pair of flexible and powerful correlators, a high (5000m) dry southern hemisphere location with a large collecting area (initially 6900 m$^2$).  It will image line emission from celestial objects with about two orders of magnitude more sensitivity than has been available, and with resolution up to two orders of magnitude better than has been provided before.
\begin{table}[ht]	\label{spectable}
\caption{Summary of ALMA Specifications}
\begin{center}
\scriptsize
\begin{tabular}{lr}
\noalign{\medskip}
\hline
\noalign{\smallskip}
Parameter & Specification \\
\hline
\noalign{\smallskip}
Number of Antennas & $>$66 \\
Antenna Diameter & 12m  \& 7m\\
Antenna Surface Precision & $<$ 25 $\mu$m rss \\
Antenna Pointing Accuracy & $<$ 0."6 rss \\
Total Collecting Area & $>$6900 m$^2$ \\
Angular Resolution & 0".015 $\lambda$ (mm) \\
Configuration Extent & 150 m to $\sim$14 km \\
Correlator Bandwidth & 16 GHz per baseline \\
Spectral Channels & 4096 per IF \\
Number of IFs & 8 \\
\hline
\hline
\end{tabular}
\end{center}
\end{table}

\section{Elements of ALMA}
\label{sec:1}

\subsection{Science Requirements }
\label{Sec:1a}

Annex B of the ALMA Bilateral Agreement set ALMA's highest level science requirements.
The highest level science requirements that have determined the ALMA parameters are the ability to: 
(1) detect spectral line emission from rotational spectral lines of the carbon monoxide molecule, 
atomic and ionized carbon in a galaxy with the properties of the Milky Way at a redshift of z=3\cite{RefC} 
in less than 24 hours of measurement, (2) image the kinematics of gas in protostars and 
protoplanetary disks around young solar type stars out to a distance of 500 light years 
\cite{RefD}, \cite{RefE}. This 
represents the distance to the nearby well-known clouds in Ophiuchus, Taurus or Corona Australis, 
and (3) provide precise images at an angular resolution better than 0.1". Here 
"precise" means that the ratio of the most intense to weakest feature in the 
image can reach 1000. This applies to sources that transit at more than 20$^{\rm o}$ 
elevation at the ALMA site\cite{RefD}. 

The key features of ALMA which will allow it to achieve these key science goals are routine milliJansky sensitivity
(as a result of the superb site, the receivers, which define the state of the art, and the large collecting
area of ALMA) and high resolution (afforded by the long baselines on the extensive site).

\subsection{Progress at the Superb Site}
\label{sec:2
}
ALMA construction has rapidly progressed.  In ALMA labs worldwide
prototypes of nearly all elements of ALMA have been tested.  These elements are now being brought together at the ALMA Test Facility in New Mexico.  There prototype integration of ALMA components into 
a functioning whole is ongoing.  In February, 2007, for example, fringes were detected from a transmitted signal external to the antennas; fringes from astronomical sources will soon be assessed.

 The site shows excellent submillimeter 
transparency--atmospheric characterization shows that 
$\tau$(490GHz)$\leq$1 for 70\% of the time during the six months 
encompassing winter.  Construction of the infrastructure necessary 
to support ALMA has reached an advanced state.  The 51 foot wide,
43km ALMA road, passable already at the ALMA groundbreaking on 2003 November 6, 
has been finished.  The 2900m altitude ALMA Camp 
sleeps and feeds ALMA personnel in its 32 bed facility while the 
Contractor Camps bed and feed  supervisors and workers (currently numbering more than 350)
with offices and recreational facilities.  ALMA personnel will move 
to the future Operations Support Facility, now in construction. The Technical
Building at the 5000m altitude Array Operations Site is complete.  John Conway, 
Mark Holdaway and collaborators have produced a
new 186-station design for the ALMA configurations, optimized for staged deployment 
of up to 64 antennas in addition to the 16 antennas of the Atacama Compact Array (ACA).  The construction of the first 
antenna pads has been finished at the 2900m facility.  Early in 2007 the 
first production antenna will arrive at the Contractor's camp for assembly 
before it moves to the project testing area in mid-2007.  As characterization
of ALMA equipment is completed and that equipment moves to the site for
scientific deployment, thoughts turn toward the scientific output of ALMA.

\subsection{Receivers}
\label{sec:3}

The four first cartridges from partners in Europe and North American have been assembled and tested in the dewar in the Front End Integration Facility at the NRAO Technology Center in Charlottesville.  Tests show that all of the preproduction cartridges are exceeding the specifications given in Table 2.  For example, for Band 6 (1.3mm), a receiver temperature of $\sim$40K has been measured in the lab.  This measurement has been verified on the sky.  Mixer/preamps for Band 6 have been sky-tested at the Submillimeter Telescope of the Arizona Radio Observatories on Mt. Graham, Arizona.  The results of these 
tests were quite impressive, with record-breaking, 
single-sideband system temperatures and exceptional 
baseline stability over wide IF bandwidths.   Typical system temperatures at elevations of about 45 degrees were around 120-140 K SSB with consistent performance across the whole frequency range of the receiver.  Image rejection was also excellent.  Although the ALMA 
image rejection specification only required 10 dB, the 
actual values were typically greater than 20 dB in the 
LSB and greater than 15 dB in the USB.  The first dewar, equipped with the four first cartridges, will be tested at the ALMA Test Facility during late Spring 2007 before being mounted on the first production antenna at the OSF later in the year.
\begin{table}[ht]	\label{senstable}
\caption{Summary of ALMA Receivers}
\begin{center}
\begin{tabular}{rcccc}
\noalign{\medskip}
\hline
\noalign{\smallskip}
  Band no. & Frequency & Receiver Noise Temperature$^a$ & Mixing Scheme & IF Bandwidth \\
           & Range (GHz)  & (K)  \\
\hline
\noalign{\smallskip}
3 & 84--116 & 37 & 2SB & 4 GHz \\
4 &125--169 & 51 & 2SB & 4 GHz\\
5$^b$ & 163--211 & 65 &2SB & 4 GHz \\
6 &211--275 & 83 & 2SB & 8 GHz \\
7 & 275--373 & 147  & 2SB & 4 GHz\\
8 & 385-500 & 98 & 2SB & 4 GHz \\
9 & 602--720 & 175 & DSB & 8 GHz \\
10$^b$ & 787--950 & 230 & DSB & 8 GHz \\
\hline
\hline
\end{tabular}
\end{center}
$^a$ Over 80\% of the band, specification.  Preproduction units tested to date have been outperforming their specifications.\\
$^b$ At first light, these bands will be available on fewer than all of the antennas in the array.\\
\end{table}

\subsection{Correlators}
\label{sec:4}
ALMA will have two correlators, one (ALMA Correlator) serving 64 elements comprised of either antennas of the main array of 12-m antennas or an array combined of these antennas with elements of the ACA antenna complement, or another correlator (ACA Correlator) which serves the 16 elements of the ACA.  To the observer, the two correlators offer nearly identical functionality and operate in a parallel fashion.
The ALMA Correlator offers a great deal of flexibility \cite{RefF}; the seventy-one supported modes of the full correlator are described in an ALMA Memo\cite{RefG}.  The observer interacts with the correlator through the Observing Tool (OT), software which generates a set of commands which execute the observation.  In general, the observer may specify a set of disjoint or overlapping spectral regions, each characterized by bandwidth (31.25 MHz to 2 GHz); each of eight 2 GHz 'baseband' inputs drives 32 tunable digital filters.  For each spectral window, the observer also specifies the central or starting frequency, the number of channels (determining spectral resolution; typically 8192 channels are available for a maximum resolution of 3.8 kHz), and the number of polarization products.  In the ALMA system, the baseband analog outputs of the antennas are digitized in a standard fashion with 3-bits at 4 Gigasamples per second; this is resampled at the correlator input with 2-bit resolution; improved sensitivity options for 4x4 bit correlation or double Nyquist modes is also available.
The temporal resolution depends on the mode chosen and may range from 16 msec to 512 msec.  Autocorrelation is also available; a 1ms time resolution can be achieved for autocorrelation data.

\begin{table}[ht]	\label{senstable}
\caption{Summary of ALMA Line Sensitivity}
\begin{center}
\begin{tabular}{rccccc}
\noalign{\medskip}
\hline
\noalign{\smallskip}
  Frequency &  B$_{max}$= 0.2 km $^a$ &   &  B$_{max}$= 14.7 km  && \\
   (GHz)  & Beamsize & $\Delta T$(K)     & Beamsize & $\Delta T$(K) &  $\Delta T$(K)  \\
               &  "  &  1 km s$^{-1}$ &  " & 1 km s$^{-1}$ & 25 km s$^{-1}$ \\
\hline
\noalign{\smallskip}
110 & 2.8 & 0.10 &   0.038 & 532 & 106  \\
140 & 2.2 & 0.10 &   0.030& 543 & 109 \\
230 & 1.3 &  0.15 &  0.018& 780 & 156 \\
345 & 0.9 & 0.23 &  0.012&  1240& 248\\
409 & 0.7&  0.34 & 0.010 & 1722& 344\\
675 & 0.4 & 0.85 & 0.006 & 4200 & 840 \\
\hline
\hline
\end{tabular}
\end{center}
$^a$ For an integration time of 60 seconds, a spectral resolution of 1 km s$^{-1}$ or 25 km s$^{-1}$, the rms brightness temperature sensitivity $\Delta T$ for an array combining all 54 12m and 12 7m antennas and a maximum baseline B$_max$ is given.  \\
$^b$ The assumed precipitable water vapor (pwv) content varies as a function of frequency. Highest frequency observations are assumed to be carried out during 'best weather' (e.g., lowest pwv) and lower frequency observations during 'worst weather'. This implies higher noise temperatures at mm wavelengths. The assumed pwv values are:  pwv = 2.3 mm for f$<$ 300 GHz; pwv = 1.2 mm for 300 $<$ f $<$ 500 GHz;  pwv = 0.5 mm for f $>$ 500 GHz..  Note that pwv=0.5 mm corresponds approximately to the 25-th percentile of the pwv distribution over time.\\
\end{table}


The tremendous sensitivity of ALMA combined with its flexible correlators will enable a wide range of science.  In the Solar System, for example, venting on small bodies can be studied in detail.  Consider Saturn's 500km diameter moon Enceladus orbiting at a distance of just under 4 Saturn radii with a period of 1.37 days.  Among the suprising features of Enceladus discovered by Cassini is the existence of fountains, probably eruptions of subsurface water.  The fountains show the spectral signature of ice particles, expected at the -201 C temperature of the surface of the satellite, but hint at the presence of liquid water deep inside the moon.  The fountains project about 50 km (about the ALMA beamsize at submillimeter frequencies) from the surface in backlit images of the limb near its south pole.  ALMA should be capable of imaging the water in these plumes, as it can easily resolve the moon's disk, providing data on these events long after Cassini ceases its observations.  ALMA will also provide time and velocity resolved images of the SO and SO$_2$ molecules emitted from the volcanoes on Io.  The ability to simultaneously image the pressure-broadened lines in planetary atmospheres in low resolution while imaging the narrow cores of the lines in high resolution mode will produce good profiles of gas distributions in planetary atmospheres.

Consider an example ALMA project, proposed by David Meier, to examine the gas and dust structure of the nearby star forming galaxy IC342 at 1.3mm wavelength.  A Hubble Wide Field Planetary Camera
 image, for example, shows the dusty center of the galaxy in an image 2'.7 across.  The ALMA primary bean at this wavelength subtends 27"; Nyquist sampling of the complementary ALMA image would require 324 pointings to construct an image cube of similar extent to the HST image.  The ALMA correlator could image J=2-1 lines of $^{12}$CO, $^{13}$CO and C$^{18}$O along with, for example, the J=10-9 line of HNCO simultaneously.  To attain a sensitivity of 10$\sigma$ for the lines, one would aim for an rms of 0.06 K, achieved in two minutes per pointing.  The entire experiment would take about eleven hours, consisting of multiple mosaics of the source.  Simultaneously, continuum observations (through binning channels free of line emission) would reach a sensitivity of 65 $\mu$Jy.  The images would provide arcsecond resolution on the galaxy's dust and gas.
 \begin{table}[ht]	\label{senstable}
\caption{Extragalactic Gas and Dust Setup}
\begin{center}
\begin{tabular}{rccccc}
\noalign{\medskip}
\hline
\noalign{\smallskip}
\hline
\noalign{\smallskip}
Line & CO & $^{13}$CO &  C$^{18}$O & HNCO & Continuum  \\
Transition & J=2-1 & J=2-1 & J=2-1 & J=10-9 &   \\
Frequency (GHz) & 230.5 & 220.4 & 219.6 & 219.8 & 4 GHz \\
Sideband & USB & LSB & LSB & LSB & USB\&LSB \\
Resolution (km s$^{-1}$) & 0.64 & 0.64 & 0.64 & 0.64 & 21. \\
Quadrant & Q1 & Q2 & Q2 & Q2 & Q3\&Q4\\
Window Bandwidth & 500MHz & 500MHz & 500MHz & 500MHz & 2$\times$2GHz\\
Channels & 1024 & 1024 & 1024 & 1024 & 2$\times$128\\
Spatial Resolution & 1", B$_{max}$=0.3km & & & & \\
\hline
\hline
\end{tabular}
\end{center}

\end{table}

A prototype suite of high-priority ALMA projects that could be carried out in about three years of full ALMA operations has been compiled in the form of the ALMA Design Reference Science Plan (DRSP).  The DRSP in comprised of more than ten dozen submissions received from a nearly equal number of astronomers.  The current version is publicly available through links at any of the ALMA websites; it is currently being expanded to include projects which exploit the enhancements made possible through Japan joining ALMA.

%
%



\begin{thebibliography}{}
%
%
\bibitem{RefC} de Breuck, C.\ 2005, In: Proceedings of the dusty and molecular universe: a prelude to Herschel and ALMA, 27-29 October 2004, Paris, France. Ed. by A. Wilson. ESA SP-577, Noordwijk, 
Netherlands: ESA Publications Division, ISBN 92-9092-855-7, 27
\bibitem{RefG}  Escoffier, R.~P., et al.\ 2006, Observational Modes Supported by the ALMA Correlator,  ALMA Memo No. 566, http://www.alma.nrao.edu/memos/html-memos/alma556/memo556.html  
\bibitem{RefF} Escoffier, R.~P., et al.\ 2007, Astronomy and Astrophysics, 462, 801 
\bibitem{RefB} McCarthy, M.~C., Gottlieb, C.~A., Gupta, H., \& Thaddeus, P.\ 2006, Ap. J. (Letters), 652, L141 
\bibitem{RefD} Richer, J.\ 2005, 
In: Proceedings of the dusty and molecular universe: a prelude to Herschel and ALMA, 
27-29 October 2004, Paris, France. Ed. by A. Wilson. ESA SP-577, Noordwijk, 
Netherlands: ESA Publications Division, ISBN 92-9092-855-7, 33
\bibitem{RefA} Wootten, A.\ 2001, ASP Conf.~Ser.~235: Science with the Atacama Large Millimeter Array, 235,  
\bibitem{RefE} Wootten, A., Mangum, 
J.~G., \& Holdaway, M.\ 2004, ASP Conf.~Ser.~324: Debris Disks and the 
Formation of Planets, 324, 277 
\end{thebibliography}


\end{document}